\documentstyle[preprint,prl,tighten,aps,epsfig]{revtex}    %

\begin{document}

\draft

\preprint{\begin{tabular}{l}
\hbox to\hsize{\mbox{ }\hfill hep--ph/9907496}\\
\hbox to\hsize{\mbox{ }\hfill KIAS-P99066}\\
\hbox to\hsize{\mbox{ }\hfill MADPH-99-1131}\\
\hbox to\hsize{\mbox{ }\hfill \today}\\
\hbox to\hsize{\mbox{ }\hfill { } }\\
\hbox to\hsize{\mbox{ }\hfill { } }\\
          \end{tabular} }

\title{ Decays of the MSSM Higgs Bosons with Explicit CP Violation}

\author{S.Y.~Choi and Jae Sik Lee}
\address{Korea Institute for Advanced Study, 207--43, Cheongryangri--dong
         Dongdaemun--gu, Seoul 130--012, Korea}

\maketitle

\vskip 2cm
\begin{abstract}
We study Higgs boson decays in the minimal supersymmetric standard model 
where the 
tree--level CP invariance of the Higgs potential is explicitly broken
by loop effects of soft CP--violating Yukawa interactions related to 
scalar quarks of the third generation.  The scalar--pseudoscalar mixing
among two neutral CP--even Higgs bosons and one CP--odd Higgs boson due 
to explicit CP violation modifies their tree--level couplings to fermions, 
to the $W^\pm$ and $Z$ bosons and to Higgs bosons themselves 
significantly. We analyze the phenomenological impact of explicit CP
violation on the branching ratios of the neutral Higgs boson decays
in detail and discuss how to directly confirm the existence of explicit 
CP violation through $\tau^+\tau^-$ and $t\bar{t}$
spin correlations in the decays of the neutral Higgs bosons into
a tau--lepton pair and a top--quark pair.
\end{abstract}

\vskip 0.4cm



\section{Introduction}
\label{sec:introduction}

Revealing the physical mechanism responsible for the breaking of the
electroweak symmetry is one of the most crucial issues in particle
physics. If the fundamental particles such as leptons, quarks and 
gauge bosons remain weakly interacting up to very high energies, the
sector in which the electroweak symmetry is broken must contain one or 
more fundamental scalar Higgs bosons with masses of the order of
the symmetry--breaking scale, However, if the Standard Model (SM) is 
embedded in a Grand Unified Theory (GUT) at high energies, the natural
scale of the electroweak symmetry breaking would be expected close to the
unification scale. Supersymmetry \cite{MSSM} provides an elegant solution 
to this hierarchy problem and furthermore leads to an excellent 
agreement between the value of the electroweak mixing angle 
$\sin^2\theta_W$ predicted by the unification of the gauge couplings 
and the experimentally measured value. 

Of course, supersymmetry must be (softly) broken to be 
phenomenologically viable. In general, this breakdown introduces a 
large number of unknown parameters, many of which can be complex \cite{MS}. 
CP--violating phases associated with sfermions of the first and, to
a lesser extent, second generations are severely constrained by bounds
on the electric dipole moments of the electron, neutron and muon.
However, there have been several suggestions \cite{KO,Kaplan,IN}
to evade these constraints  without suppressing the CP--violating phases.
One option is to make the first two generations of scalar fermions
rather heavy so that one--loop EDM constraints are automatically evaded. 
As a matter of fact one can consider so--called effective
SUSY models \cite{Kaplan} where de-couplings of the first and second 
generation sfermions are invoked to solve the SUSY FCNC and CP problems 
without spoiling the naturalness condition.
Another possibility is to arrange for partial cancellations among 
various contributions to the electron and neutron EDM's \cite{IN}.

Following the suggestions that the phases do not have to be suppressed,
many important works on the effects due to the CP phases in the minimal
supersymmetric standard model (MSSM) have been already reported; 
the effects are very significant in extracting the parameters in the 
SUSY Lagrangian from experimental data \cite{SYCHOI}, 
estimating dark matter densities and scattering cross sections and 
Higgs boson mass limits \cite{BK,FO,PW}, CP violation in the $B$ and $K$ 
systems \cite{Ko}, and so on. In particular, it has been found \cite{PW} 
that the Higgs--sector CP violation induced via loop corrections of
soft CP--violating Yukawa interactions  may drastically modify the 
couplings of the light neutral Higgs boson to the gauge bosons. 
As a result, the current experimental lower bound on the lightest Higgs 
boson mass may be dramatically relaxed up to a 60--GeV level in the 
presence of large CP violation in the Higgs sector of the MSSM. 

Since the experimental
observation of scalar Higgs particles and the detailed confirmation
of their fundamental properties are crucial for our present understanding
of the mechanism of the electroweak symmetry breaking,  
a very precise prediction of the production cross sections and of the 
branching ratios for the main decay channels is mandatory. 
So, several recent works \cite{PW} have made the detailed predictions 
of the Higgs mass spectrum and their production cross sections in the MSSM 
with explicit CP violation. Along with the mass spectrum and cross 
sections, it is also very important to estimate the branching ratios of 
the main Higgs boson decays precisely. This indispensable requirement 
leads to our   systematic study of the Higgs boson decays in the 
MSSM \cite{Spira} with explicit CP violation, where both Higgs--boson 
masses and their couplings to fermions, gauge bosons are significantly 
affected by loop corrections of soft CP--violating Yukawa interactions
to the Higgs sector.
We provide a detailed quantitative estimate for the branching fractions
of all possible two--body decay modes of three neutral Higgs bosons
and investigate the possibility of measuring CP violation
through the fermion spin correlations in the neutral Higgs decays
into a tau--lepton pair and/or a top--quark pair.

The organization of the paper is as follows. In Section \ref{sec:CP
violation} we give a brief review for the explicit CP violation
in the Higgs sector due to the significant contributions of  radiative 
corrections from trilinear Yukawa couplings of the Higgs fields to scalar
top and scalar bottom quarks. This review is mainly based on the work 
by Pilaftsis and Wagner \cite{PW}. The parts involving other supersymmetric
particles such as charginos and neutralinos  can contribute to the effective 
Higgs potential. But, the chargino and neutralino contributions 
are degraded in the present analysis with their masses taken to be
of the order of a large SUSY breaking scale $M_{\rm SUSY}$. 
Then, we introduce
the general mass matrix dictating the mixing among three neutral Higgs
states and the resulting Lagrangian describing the couplings of the
Higgs bosons to fermions, gauge bosons, and Higgs bosons themselves. 
In Section \ref{sec:decay} we discuss the effects of the CP 
phases on (almost) all the two--body Higgs boson decays in detail, 
which will give an important phenomenological impact on Higgs--boson 
searches. Section \ref{sec:measure} is devoted to investigating the 
possibility of detecting CP--violating effects in the neutral Higgs 
boson decays
into tau--lepton pair or top--quark pair in which the subsequent decays
of tau leptons and top quarks enable one to extract the information on the
spin of the parent fermions; tau leptons and top quarks. 
Finally, Section \ref{sec:conclude} summarizes our findings and concludes.

\section{CP Violation in the MSSM Higgs Sector}
\label{sec:CP violation}

The MSSM introduces several new parameters in the theory that are absent
from the SM and could, in principle, possess many CP--violating phases. 
Specifically, the new CP phases may come from the following parameters: 
(i) the higgsino mass parameter $\mu$, which involves the bilinear mixing 
of the two Higgs chiral superfields in the superpotential; (ii) the soft 
SUSY--breaking gaugino masses $M_a$ ($a=1,2,3$), where the index $a$ 
stands for the gauge groups U(1)$_Y$, SU(2)$_L$ and SU(3)$_C$, 
respectively; (iii) the soft bilinear Higgs mixing masses $m^2_{12}$, 
which is sometimes denoted as $B\mu$ in the literature; (iv) the soft 
trilinear Yukawa couplings $A_f$ of the Higgs particles to scalar 
fermions; and (v) the flavor mixing elements of the sfermions mass matrices. 
If the universality condition is imposed on all gaugino masses 
at the unification scale $M_X$, the gaugino masses $M_a$ have a common 
phase, and if the diagonal boundary conditions are added to the
universality condition for the sfermion mass matrices at the GUT
scale, the flavor mixing elements of the sfermions mass matrices vanish 
and the different trilinear couplings $A_f$ are all equal, i.e. $A_f=A$. 

The conformal--invariant part of the MSSM Lagrangian has two 
global U(1) symmetries; the U(1)$_Q$ Peccei--Quinn symmetry
and the U(1)$_R$ symmetry acting on the Grassmann--valued coordinates.
As a consequence, not all CP--violating phases of the four complex parameters 
$\{\mu,m^2_{12},M_a,A\}$ turn out to be physical, i.e. two phases may be
removed by redefining the fields accordingly \cite{DGH}. 
Employing the two global 
symmetries, one of the Higgs doublets and the gaugino
fields can be rephased such that $M_a$ and $m^2_{12}$ become real.
In this case, arg($\mu$) and arg($A$) are the only physical
CP--violating phases in the low--energy MSSM supplemented by universal 
boundary conditions at the GUT scale.
Denoting the scalar components of the Higgs doublets $H_1$ and $H_2$ by
$H_1=-i\tau_2 \Phi^*_1$ ($\tau_2$ is the usual Pauli matrix) and 
$H_2=\Phi_2$, the most general CP--violating 
Higgs potential of the MSSM can be conveniently described by the effective
Lagrangian
\begin{eqnarray}
{\cal L}_V&=&\mu^2_1(\Phi^\dagger_1\Phi_1)
        +\mu^2_2(\Phi^\dagger_2\Phi_2)
        +m^2_{12}(\Phi^\dagger_1\Phi_2)
        +m^{*2}_{12}(\Phi^\dagger_2\Phi_1)\nonumber\\
      &&+\lambda_1(\Phi^\dagger_1\Phi_1)^2
        +\lambda_2(\Phi^\dagger_2\Phi_2)^2
        +\lambda_3(\Phi^\dagger_1\Phi_1)(\Phi^\dagger_2\Phi_2)
        +\lambda_4(\Phi^\dagger_1\Phi_2)(\Phi^\dagger_2\Phi_1)\nonumber\\
      &&+\lambda_5(\Phi^\dagger_1\Phi_2)^2
        +\lambda^*_5(\Phi^\dagger_2\Phi_1)^2
        +\lambda_6(\Phi^\dagger_1\Phi_1)(\Phi^\dagger_1\Phi_2)
        +\lambda^*_6(\Phi^\dagger_1\Phi_1)(\Phi^\dagger_2\Phi_1)\nonumber\\
      &&+\lambda_7(\Phi^\dagger_2\Phi_2)(\Phi^\dagger_1\Phi_2)
        +\lambda^*_7(\Phi^\dagger_2\Phi_2)(\Phi^\dagger_2\Phi_1)\,.
\label{eq:Higgs potential}
\end{eqnarray}
In the Born approximation, the quartic couplings $\lambda_{1,2,3,4}$ are 
solely determined by the gauge couplings and $\lambda_{5,6,7}$ are zero. 
However, beyond the Born approximation, the quartic couplings 
$\lambda_{5,6,7}$ receive significant radiative corrections from 
trilinear Yukawa couplings of the Higgs fields to scalar--top and 
scalar--bottom quarks. These parameters are in general complex and so 
lead to CP violation in the Higgs sector through radiative corrections. 
The explicit form of the couplings with the radiative corrections included
can be found in Refs.~\cite{PW,HH}.

It is necessary to determine the ground state of the Higgs potential to 
obtain physical Higgs states and their self--interactions. To this end
we introduce the linear decompositions of the Higgs fields
\begin{eqnarray}
\Phi_1=\left(\begin{array}{cc}
             \phi^+_1 \\
	     \frac{1}{\sqrt{2}}(v_1+\phi_1+ia_1)
	     \end{array}\right)\,, \qquad
\Phi_2={\rm e}^{i\xi}\left(\begin{array}{cc}
             \phi^+_2 \\
	     \frac{1}{\sqrt{2}}(v_2+\phi_2+ia_2)
	     \end{array}\right)\,,
\end{eqnarray}
with $v_1$ and $v_2$ the moduli of the vacuum expectation values (VEVs) 
of the Higgs doublets and $\xi$ their relative induced CP--violating phase. 
These VEVs and the 
relative phase can be determined by the minimization conditions on 
${\cal L}_V$, which can be efficiently performed by the so--called tadpole 
renormalization techniques \cite{PW}. It is always guaranteed that one 
combination of the CP--odd  Higgs fields $a_1$ and $a_2$ ($G^0=\cos\beta a_1
-\sin\beta a_2$) defines a flat direction in the Higgs potential and 
so it is absorbed as the longitudinal component of the $Z$ boson. 
[Here, $\sin\beta=v_2/\sqrt{v^2_1+v^2_2}$ and 
$\cos\beta=v_1/\sqrt{v^2_1+v^2_2}$.] As a result, there exist one charged 
Higgs state and three neutral Higgs 
states that are mixed in the presence of CP violation in the Higgs 
sector. Denoting the remaining CP--odd state $a=\-\sin\beta a_1
+\cos\beta a_2$, the $3\times 3$ neutral Higgs--boson mass matrix 
describing the mixing between CP--even and CP--odd fields can be 
decomposed into four parts in the weak basis $(a,\phi_1,\phi_2)$ :
\begin{eqnarray}
{\cal M}^2_0=\left(\begin{array}{cc}
                   {\cal M}^2_P     &   {\cal M}^2_{PS} \\
		   {\cal M}^2_{SP}  &   {\cal M}^2_S 
		   \end{array}\right)\,,
\end{eqnarray}
where ${\cal M}^2_P$ and ${\cal M}^2_S$ describe the CP--preserving
transitions $a\rightarrow a$ and $(\phi_1,\phi_2)\rightarrow 
(\phi_1,\phi_2)$, respectively, and ${\cal M}^2_{PS}=({\cal M}^2_{SP})^T$ 
contains the CP--violating mixings $a\leftrightarrow (\phi_1,\phi_2)$. 
The analytic form of the sub--matrices is given by
\begin{eqnarray}
&&{\cal M}^2_P=m^2_a=\frac{1}{s_\beta c_\beta}\left\{
            {\cal R}(m^2_{12}{\rm e}^{i\xi})
            +v^2\left[2{\cal R}(\lambda_5{\rm e}^{2i\xi})s_\beta c_\beta
	    +\frac{1}{2}{\cal R}(\lambda_6{\rm e}^{i\xi})c^2_\beta
	    +\frac{1}{2}{\cal R}(\lambda_7{\rm e}^{i\xi})s^2_\beta\right]
	    \right\}\,, \nonumber\\
&&{\cal M}^2_{SP}=v^2\left(\begin{array}{c}
             {\cal I}(\lambda_5{\rm e}^{2i\xi})s_\beta
	    +{\cal I}(\lambda_6{\rm e}^{i\xi})c_\beta\\
             {\cal I}(\lambda_5{\rm e}^{2i\xi})c_\beta
	    +{\cal I}(\lambda_7{\rm e}^{i\xi})s_\beta
	    \end{array}\right)\,, \nonumber\\
&&{\cal M}^2_S=m^2_a\left(\begin{array}{cc}
             s^2_\beta        & -s_\beta c_\beta\\
            -s_\beta c_\beta & c^2_\beta
	    \end{array}\right)\nonumber\\
&& -\left(\begin{array}{cc}
      2\lambda_1 c^2_\beta+2{\cal R}(\lambda_5{\rm e}^{2i\xi})s^2_\beta
 +2{\cal R}(\lambda_6{\rm e}^{i\xi})s_\beta c_\beta &
 \lambda_{34}s_\beta c_\beta+{\cal R}(\lambda_6{\rm e}^{i\xi})c^2_\beta
 +{\cal R}(\lambda_7{\rm e}^{i\xi})s^2_\beta \\ 
 \lambda_{34}s_\beta c_\beta+{\cal R}(\lambda_6{\rm e}^{i\xi})c^2_\beta
 +{\cal R}(\lambda_7{\rm e}^{i\xi})s^2_\beta & 
       2\lambda_2 s^2_\beta+2{\cal R}(\lambda_5{\rm e}^{2i\xi})c^2_\beta
 +2{\cal R}(\lambda_7{\rm e}^{i\xi})s_\beta c_\beta 
 \end{array}\right)\,,
\end{eqnarray}
where $c_\beta=\cos\beta$ and $s_\beta=\sin\beta$ and
the mass squared $m^2_a$ is given by
\begin{eqnarray}
m^2_a=\frac{1}{s_\beta c_\beta}\left\{{\cal R}(m^2_{12}{\rm e}^{i\xi})
     +v^2\left[2{\cal R}(\lambda_5{\rm e}^{2i\xi})s_\beta c_\beta
              +\frac{1}{2}{\cal R}(\lambda_6{\rm e}^{i\xi})c^2_\beta
              +\frac{1}{2}{\cal R}(\lambda_7{\rm e}^{i\xi})s^2_\beta
         \right]\right\}.
\end{eqnarray}
Correspondingly, the charged Higgs-boson mass is given by
\begin{eqnarray}
m^2_{H^\pm}=m^2_a+\frac{1}{2}\lambda_4 v^2
           -{\cal R}(\lambda_5{\rm e}^{2i\xi}) v^2\,.
\end{eqnarray}
Taking this very last relation between $m_{H^\pm}$ and $m_a$ into account,
we can express the neutral
Higgs--boson masses as functions of $m_{H^\pm}$, $\mu$, $A_t$, $A_b$,
a common SUSY scale $M_{\rm SUSY}$, $\tan\beta$ and the physical
phase $\xi$. However, with the chargino and neutralino contributions
neglected, the radiatively induced phase $\xi$ can be absorbed into the 
definition of the $\mu$ parameter.
Clearly, the CP--even and CP--odd states mix unless all of the
imaginary parts of the parameters $\lambda_{5,6,7}$ vanish. Since
the Higgs---boson mass matrix ${\cal M}^2_0$ describing the 
scalar--pseudoscalar mixing is 
symmetric, we can diagonalize it by means of an orthogonal rotation $O$;
$O^T{\cal M}^2_T O={\rm diag}(m^2_{H_3},m^2_{H_2},m^2_{H_1})$
with the ordering of masses $m_{H_1}\leq m_{H_2}\leq m_{H_3}$.
The neutral Higgs--boson mixing affects the couplings of the Higgs fields 
to fermions, gauge bosons, and Higgs fields themselves as shown in the
following. On the other hand, there could exist CP--violating vertex 
correctios, but they have been shown to be rather small \cite{Vertex}
so that those effects are not included in the present work.

Firstly, the interactions of the neutral Higgs fields with SM fermions are 
described by the Lagrangian
\begin{eqnarray}
&& {\cal L}_{H\bar{f}f}=-\frac{g}{2m_W}\frac{m_f}{R^f_\beta}\, 
    \bar{f}\left[v^i_f-i\bar{R}^f_\beta a^i_f\gamma_5\right]f\, H_i\,, \\
&& R^f_\beta =\left\{\begin{array}{c} c_\beta   \\ 
   s_\beta\end{array}\right.\,,
         \  \
   \bar{R}^f_\beta =\left\{\begin{array}{c} s_\beta\\ 
   c_\beta\end{array}\right.\,,
         \  \
   v^i_f =\left\{\begin{array}{c} O_{2,4-i}\\ O_{3,4-i} 
          \end{array}\right.\,,
         \  \
   a^i_f =\left\{\begin{array}{c} O_{1,4-i}\\ O_{1,4-i} 
          \end{array}\right.\,,
         \  \
   \begin{array}{l} {\rm for}\ \ f=(l:d) \\ {\rm for}\ \ f=(u)  
         \end{array}\,.
\end{eqnarray}
Obviously, the Higgs--fermion--fermion couplings are significant for the 
third--generation fermions, $t$, $b$ and $\tau$ because of their relatively
large Yukawa couplings. We readily see that the 
effect of CP--violating Higgs mixing is to induce a simultaneous coupling 
of $H_i$ ($i=1,2,3$) to CP--even and CP--odd fermion bilinears $\bar{f}f$
and $\bar{f} i\gamma_5 f$ \cite{DM}. This can lead to a sizable phenomenon 
of CP 
violation in the Higgs decays into polarized top-quark or tau-lepton 
pairs \cite{CKCK}, which will constitute the topic of  
Section \ref{sec:measure}.

Secondly, the couplings of the Higgs fields to $W$ and $Z$ bosons may be 
read off by the Lagrangian
\begin{eqnarray}
&&{\cal L}_{HVV}=gm_W(c_\beta O_{2,4-i}+s_\beta O_{3,4-i})
    \, H_i\left[W^+_\mu W^{-\mu}+\frac{1}{2c^2_W} Z_\mu Z^\mu\right]\,,\\
&&{\cal L}_{HH^\pm H^\mp}=\frac{g}{2}(c_\beta O_{3,4-i}-s_\beta O_{2,4-i}
     +i O_{1,4-i})\, W^{+\mu} 
    \left( H_i i\stackrel{\leftrightarrow}{\partial_\mu}
     H^-\right) +{\rm h.c.}\,, \\
&&{\cal L}_{HHZ}=\frac{g}{4c_W}
     \bigg[O_{1,4-i}(c_\beta O_{3,4-j}-s_\beta O_{1,4-j})
          -O_{1,4-j}(c_\beta O_{3,4-i}-s_\beta O_{1,4-i})\bigg]
	  Z^\mu \left( H_i i\stackrel{\leftrightarrow}{\partial_\mu} H_j\right)\,.
\end{eqnarray}
Note that the $Z$ boson can only couple to two different Higgs particles. 
The reason is that Bose symmetry forbids any antisymmetric derivative 
coupling of a vector particle to two identical real scalar fields.
On the other hand, the orthogonality of the mixing matrix $O$ and the 
assumption ${\rm det}O=1$, which we can take without loss of any 
generality, leads to the following important relation between the 
couplings of the neutral Higgs bosons to the gauge bosons:
\begin{eqnarray}
g_k=\epsilon_{ijk} g_{ij},
\label{eq:relation}
\end{eqnarray}
with the definitions; $g_i= c_\beta O_{2i}+s_\beta O_{3i}$ and
$g_{ij}= O_{1i}(c_\beta O_{3j}-s_\beta O_{2j})
        -O_{1j}(c_\beta O_{3i}-s_\beta O_{2i})$.
One intermediate result from the relation and the unitarity constraint
is that the knowledge of two $g_i$ is sufficient to determine the whole
set of couplings of the neutral Higgs to the gauge bosons \cite{MPGG}.
As will be seen in Sec.~III C, the above relation leads to a close 
correlations between ${\cal B}(H_2 \rightarrow H_1 Z)$ and 
${\cal B}(H_3 \rightarrow V V)$ and it also leads to a similar
close relation between  ${\cal B}(H_3 \rightarrow V V)$ and
${\cal B}(H_2 \rightarrow H_1 Z)$.

Thirdly, the trilinear Higgs self--couplings also are affected by the 
radiative corrections. Their relevant interactions can be read off by 
the Lagrangian
\begin{eqnarray}
{\cal L}_{HHH}=v\bigg[A_{ijk} H_i H_j H_k + B_i H_i H^+ H^-\bigg]\,
\end{eqnarray}
where the totally symmetric $A_{ijk}$ and the coefficients $B_i$ are 
given by
\begin{eqnarray}
A_{ijk}=\sum_{\alpha\beta\gamma=1,2,3}O_{\alpha,4-i}
        O_{\beta,4-j} O_{\gamma, 4-k} a_{\alpha\beta\gamma} \,, \qquad
B_i=\sum_{\alpha=1,2,3} O_{\alpha, 4-i} b_\alpha\,,
\end{eqnarray}
where the totally-symmetric coefficients $a_{\alpha\beta\gamma}$ and the 
coefficients $b_i$ can be directly obtained from taking every possible 
combination of third--order derivatives of the general Higgs potential 
(\ref{eq:Higgs potential}) with respect to the fields $\{a,\phi_1,\phi_2,
H^\pm\}$.
The coefficients, of which the form is presented in the Appendix, are 
functions of the quartic couplings $\lambda_i$ ($i=1$ to $7$).
As can be checked in the Lagrangian, the radiative corrections generate 
various new interaction vertices among the neutral Higgs fields and the 
charged Higgs fields, which are absent in the CP--invariant MSSM.

Finally, we emphasize that the size of CP violation due 
to radiative corrections to the Higgs potential is characterized by a 
dimensionless parameter $\eta_{CP}$
\begin{eqnarray}
\eta_{_{CP}}=\frac{m^4_f}{v^4}\left(\frac{|\mu||A_f|}{32\pi^2 M^2_{\rm SUSY}}
          \right)\sin\Phi\,,
\end{eqnarray}
where $\Phi={\rm arg}(A_f\mu)+\xi$, i.e. the sum of three CP--violating
phases. So, for $|\mu|$ and $|A_f|$ values larger than the SUSY--breaking 
scale $M_{\rm SUSY}$, the CP--violating effects can be significant.

\section{Higgs Boson Decays}
\label{sec:decay}

\subsection{ Higgs--boson masses}

Higgs-boson masses are not determined a priori within the theory and their
decay patterns depend strongly on the masses as well as several SUSY 
parameters. So, in order to determine the possible decay modes and 
branching ratios, it is necessary to investigate the change of the mass 
spectrum with respect to the SUSY parameters. 
As explained in Section~\ref{sec:CP violation}, the mass spectrum is 
determined by nine real quantities 
$\tan\beta$, $m_{H^\pm}$, $|\mu|$, $|A_t|$, $|A_b|$, $\Phi_\mu$,
$\Phi_{A_t}$, $\Phi_{A_b}$, and $M_{\rm SUSY}$.  For simplicity, we 
assume in our numerical analysis for the following dimensionful parameters
\begin{eqnarray}
|A_t| = |A_b| = 1\,{\rm TeV}\,, \  \ |\mu| = 2\,{\rm TeV}\,, \  \
M_{\rm SUSY}=0.5\,{\rm TeV}\,, 
\label{eq:parameter}
\end{eqnarray}
while we treat $m_{H^\pm}$ as a dimensionful free parameter.
Noticing that the effects of the CP phases appear with the unique
combination $\Phi={\rm arg}(A_f\mu)+\xi$ and using the freedom of field 
redefinitions guaranteed only if chargino and neutralino contributions
are neglected, we take for the phases
\begin{eqnarray}
\Phi_\mu + \xi =0\,, \  \ \Phi_{A_t}=\Phi_{A_b}\equiv \Phi\,. 
\end{eqnarray}
Then, we are left with two dimensionless free parameters
$\tan\beta$ and $\Phi$. So, eventually, all the mass spectrums and 
couplings are determined completely by three parameters 
$\{\tan\beta, \Phi, m_{H^\pm}\}$.

Figure~1 shows the masses of three neutral Higgs bosons and one charged 
Higgs boson with respect to the charged Higgs mass $m_{H^\pm}$
for different values of the CP phase $\Phi$ with the fixed values for 
other SUSY parameters in Eq.~(\ref{eq:parameter}). 
The value of $\tan\beta$ \cite{tanb} is taken to be 3 and an artificial 
experimental condition 
$m_{H_1}\geq 70$ GeV is imposed. The lower solid line is 
for the lightest Higgs boson $H_1$, the upper solid line for the heaviest
Higgs boson $H_3$, and the dashed line for the intermediate Higgs boson 
$H_2$. Comparing the mass spectrums of the whole frames, we find the 
following interesting features for the Higgs boson masses:
\begin{itemize}
\item The mass of the lightest Higgs boson $H_1$ is always less than 130 
      GeV irrespective of the value of $\Phi$.
\item The intermediate Higgs--boson mass is very sensitive to the 
      CP--violating phase $\Phi$ 
      and becomes very close to the lightest Higgs boson mass around 
      $\Phi\sim 70^0$. 
\item The heaviest Higgs--boson is almost degenerate with the charged 
      Higgs boson except the region with a small charged Higgs boson mass.
\item Except the lightest Higgs boson mass, the other Higgs boson 
      masses increase with the charged Higgs mass, which is treated 
      as a free parameter.
\end{itemize}
One can also notice that the the minimum values of the allowed heavier
Higgs masses under the constraint $m_{H_1}\geq 70$ GeV depend strongly 
on the CP phase. These features clearly suggest that while the decay 
pattern of the lightest neutral Higgs boson might be rather insensitive 
to the CP phase and the charged Higgs mass, the decay patterns of the other 
neutral Higgs bosons and the charged Higgs boson itself are very 
sensitive to them.

\subsection{Higgs decay branching ratios}

In most cases, the most important decay channels of the Higgs bosons are 
two--body decays to the heaviest particles because the Higgs couplings
to the SM particles are proportional to their masses. Sometimes, 
because of the coupling enhancement below--threshold three--body decays
of the Higgs bosons may also be important \cite{DKZ}. Nevertheless, 
in this work we concentrate on only two--body decay channels. 
On the other hand, if supersymmetric particles are light enough, 
Higgs bosons can decay into supersymmetric particles. However, since we 
have assumed a common SUSY scale of the order of 1 TeV, decays to 
sfermions, neutalinos, and charginos are not expected to play an
important role in the Higgs mass 
range of a few hundred GeV which we are analyzing.

Let us now present the explicit form of each Higgs boson decay width. 
First of all, the decay width for the Higgs boson decays into a 
fermion--pair is given by
\begin{eqnarray}
\Gamma(H_i\rightarrow f\bar{f})=\frac{N_c G_F}{4\sqrt{2}\pi}
     m^2_f m_{H_i}\, \frac{\beta_f}{(R^f_\beta)^2}
     \left[ (\beta_f v^i_f)^2+ (\bar{R}^f_\beta a^i_f)^2\right]\,,
\end{eqnarray}
where $N_c=3$ and $1$ for quarks and leptons, and $\beta_f$ is the
fermion velocity $\sqrt{1-4m^2_f/m^2_{H_i}}$ in the rest frame of the
Higgs boson $H_i$. Among these fermionic decay channels, the 
bottom--quark and tau--lepton modes give most dominant contributions 
to the decay width of the lightest Higgs boson due to their relatively 
large Yukawa couplings. Numerically, the branching ratio of the 
bottom--quark channel is almost 90\% and the tau--lepton mode is 10\% in the
almost whole range of the allowed lightest Higgs boson mass. 
Moreover, if the masses of the heavier Higgs bosons are smaller than the 
threshold of two $W$'s, they will be the main decay channels for the 
heavy states as well.  As stated in the expression, the fermionic decay 
width is strongly dependent on the fermion mass so that the exact 
determination of the mass is required for the precise estimate of the 
branching ratio. Keeping this point in mind, we simply take $m_b=4.25$ GeV 
and $m_c=1.25$ GeV for our numerical analysis.

Secondly, the decay widths of the Higgs bosons into a gauge boson pair 
can be obtained by the Lagrangian for the Higgs couplings to the gauge 
bosons. The analytic form of those decay widths can be summarized as 
follows:
\begin{eqnarray}
\Gamma(H_i\rightarrow VV)=\frac{G_F m^3_{H_i}}{16\sqrt{2}\pi}\delta_V\beta_V
   \bigg|c_\beta O_{2,4-i}+s_\beta O_{3,4-i}\bigg|^2 
   \left(1-4\kappa^i_V+ 12\kappa^{i2}_V\right)\,,
\end{eqnarray}
where $V=W^\pm$ or $Z$, $\delta_V$ is 2 for the $W$ boson and 1 for the 
$Z$ boson,
$\beta_V$ is the velocity of the gauge boson in the rest frame of the
Higgs boson, and $\kappa_V^i=m_V^2/m_{H_i}^2$. 
Note that two decay widths are of the same form and for 
$m_{H_i}\gg m_Z$  the ratio of $W^+W^-$ to $ZZ$ rates approaches $2$.
Clearly, since the decay widths are proportional to the third power of
the Higgs boson mass itself, these decay channels are expected to 
contribute dominantly to the decay widths of the heavy Higgs bosons.
On the contrary, the partial width of the decay $H_1\rightarrow VV$ is 
strongly suppressed
or vanishing due to kinematics, thus not playing a dominant role even
if the decay channel with one off--shell gauge boson is considered.
In the Born approximation, the intermediate Higgs boson is the 
pseudoscalar boson so that the decay $H_2\rightarrow ZZ$ is not allowed. 
However, beyond the Born approximation, this mass eigenstate will have a 
large portion of the light or heavy scalar Higgs states and as a result 
there will be a large enhancement for the branching ratio for non--trivial
values of the CP phase $\Phi$ in this specific decay channel unique to
the CP--noninvariant theory. 

Thirdly, the heavy Higgs--boson decays into a light
Higgs boson and a gauge boson can be classified into the charged decays
$H_i\rightarrow H^\pm W^\mp$ and the neutral decays $H_i\rightarrow H_j Z$.
The width for the decay $H_i\rightarrow H^\pm W^\mp$ is given by
\begin{eqnarray}
\Gamma(H_i\rightarrow H^\pm W^\mp) = \frac{G_Fm^3_{H_i}}{8\sqrt{2}\pi}
     \,\bigg|c_\beta O_{3,4-i} - s_\beta O_{2,4-i}+i O_{1,4-i}\bigg|^2
     \,\lambda^{3/2}\left(1,\frac{m^2_W}{m^2_{H_i}},
                            \frac{m^2_{H^\pm}}{m^2_{H_i}}\right)\,,
\end{eqnarray}
where $\lambda$ is the two--body phase space function; 
$\lambda(x,y,z) = x^2+y^2+z^2-2xy-2yz-2zx$.
Similarly, the width for the decay $H_i\rightarrow H_j Z$ is given by
\begin{eqnarray}
\Gamma(H_i\rightarrow H_j Z) = \frac{G_Fm^3_{H_i}}{8\sqrt{2}\pi}
     \bigg|O_{1,4-i}(c_\beta O_{3,4-j} - s_\beta O_{2,4-i})
     - (i\leftrightarrow j)\bigg|^2
     \lambda^{3/2}\left(1,\frac{m^2_{H_j}}{m^2_{H_i}},
                            \frac{m^2_Z}{m^2_{H_i}}\right)\,.
\end{eqnarray}
The branching ratios for the two--body decays $H_2\rightarrow H_1 Z$
can be sizable in specific regions of the SUSY parameter space, 
especially, for small values of $\tan\beta$ and below the $tt$ thresholds. 

Fourthly, the heavy Higgs bosons can decay into a pair of lighter scalars,
$H_i\rightarrow H_j H_k$ and into a charged Higgs--boson pair
$H_i\rightarrow H^+ H^-$, if they are kinematically allowed. The width 
of the latter decay channel is given by
\begin{eqnarray}
\Gamma(H_i\rightarrow H^+ H^-) = \frac{v^2|B_i|^2}{16\pi m_{H_i}}
   \beta_{H^\pm}\,,
\end{eqnarray}
with $\beta_{H^\pm}$ the velocity of the charged Higgs boson in the
rest frame of the decaying neutral Higgs boson $H_i$ and $v^2=v_1^2+v_2^2$. 
On the other hand,
the width for the former decay channel $H_i\rightarrow H_j H_k$ is given by
\begin{eqnarray}
\Gamma(H_i\rightarrow H_j H_k)=\frac{9\eta_{jk}}{8\pi m_{H_i}}
       v^2|A_{ijk}|^2\beta_{ijk}\,,
\end{eqnarray}
where $\eta_{jk}$ is 2 for $j\neq k$ and 1 for $j=k$, and $\beta_{ijk}=
\lambda^{1/2}(1, m^2_{H_j}/m^2_{H_i},m^2_{H_k}/m^2_{H_i})$.

Finally, the decays of the Higgs bosons into two gluons \cite{EGN}, which
occur through quark one--loop diagrams may be sizable.  As a matter 
of fact, the two--gluon partial width of the Higgs bosons is important 
because it is the main production mechanism for the Higgs boson
production at the $pp$ collider LHC. Moreover, since gluons 
couple to Higgs bosons via heavy particle loops, the two--gluon
widths are sensitive to heavy particle masses, standard and also 
supersymmetric,
well above the Higgs masses themselves. It is a little involved but
straightforward to calculate the width for the Higgs boson decay into
two gluons, which can be expressed in the following compact form
\begin{eqnarray}
\Gamma(H_i\rightarrow gg)=\frac{G_F\alpha^2_S m^3_{H_i}}{16\sqrt{2}\pi^3}
      \sum_f\left[\left(\frac{v^i_f}{R^f_\beta}\right)^2 |F(\tau)|^2
                 +\left(\frac{\bar{R}^f_\beta a^i_f}{R^f_\beta}\right)^2 
                  |G(\tau)|^2\right]\,,
\end{eqnarray}
where $\tau=m^2_{H_i}/4m^2_f$, $f$ runs for all colored fermions,
and the functions $F(\tau)$ and $G(\tau)$ are defined by
\begin{eqnarray}
F(\tau)=\tau^{-1}\left[1+(1-\tau^{-1})f(\tau)\right]\,, \qquad
G(\tau)=\tau^{-1}f(\tau)\,, \qquad
\end{eqnarray}
with 
\begin{eqnarray}
f(\tau)=\left\{\begin{array}{ll}
      {\rm arcsin}^2\sqrt{\tau} & {\rm for}\, \tau\leq 1\,, \\
      -\frac{1}{4}\left[\log 
             \frac{\sqrt{\tau}+\sqrt{\tau-1}}{\sqrt{\tau}-\sqrt{\tau-1}}
            -i\pi\right]^2  & {\rm for}\, \tau > 1\,.
         \end{array}\right.
\end{eqnarray}
The QCD radiative corrections \cite{SDGZ} to the two--gluon channel may 
be sizable 
and are built up by the exchange of virtual gluons, gluon radiation
from the internal quark loop and the splitting of a gluon into two 
unresolved gluons or quark--antiquark pair. Although they are large 
enough to nearly double the partial width, we neglect the corrections
in our numerical analyses because after all the two--gluon channel 
remains as a sub--dominant decay channel for all the neutral Higgs
bosons. We note in passing that the two--photon channel and the other
similar one--loop decay channels have much smaller branching ratios
so that they are not included in our present discussion.

\subsection{Numerical results on the branching ratios}

In this section, we present a detailed numerical analysis of
the total decay widths of each neutral
Higgs boson, the branching ratios of each decay mode of the Higgs boson
with respect to each Higgs--boson mass for various 
values of the CP phase $\Phi$. 

Figure~2 shows the total decay widths for the three neutral Higgs bosons
with respect to each Higgs--boson mass; the left (right) solid line 
is for the lightest (heaviest) Higgs boson and the dashed line for the 
intermediate Higgs boson. The decay width
$\Gamma(H_1)$ is almost independent of the CP phase, but the other two
total widths $\Gamma(H_{2,3})$ are strongly dependent on the CP phase
at the lower ends of their allowed masses.
The widths are rather small in size so that they
may not be very hard to be determined at $e^+e^-$ colliders, but they
may be detectable at a $\mu^+\mu^-$ collider with a very precise
energy calibration. 

Figure~3 shows the partial branching ratios of the lightest Higgs boson
decays with respect to the mass $m_{H_1}$ for several 
values of the CP phase $\Phi$. Since the lightest Higgs boson mass is
very strongly restricted to be less than 130 GeV, the 
main decay channels are into $b\bar{b}$, $\tau^+\tau^-$, $gg$, and 
$c\bar{c}$. What can be immediately noted in Figure~3 is that
the pattern of the branching ratios is (almost) independent of the
CP phase. We have confirmed quantitatively that this feature is not much 
changed even if the possible off--shell decay channel $H_1\rightarrow W^{+*} 
W^-$ is included. Numerically, we find that the CP phase $\Phi$ tends 
to reduce the contribution from the off--shell decay mode. 

On the contrary,
as shown in Figure~4 the partial branching ratios for the $H_2$ decay
channels with respect to the mass $m_{H_2}$ are very sensitive to the 
CP phase $\Phi$. The upper solid line in each figure frame
denotes the sum of the branching ratios of the fermionic decay modes.
The dot--dashed line is for the decay channel $H_2\rightarrow H_1 Z$,
two dotted lines for the channels $H_2\rightarrow W^+W^-$ (upper line)
and $ZZ$ (lower line),
respectively, and the dashed line for the decay channel $H_2\rightarrow 
H_1 H_1$. Finally, the lower solid line is for the two--gluon channel.
Concerning those branching ratios, there are several interesting points 
worthwhile to be mentioned.
\begin{itemize}
\item Through the whole range of the allowed $H_2$ mass, the fermionic 
      decay channel remains as one of the dominant channels. However,
      the decay channel $H_2\rightarrow H_1 Z$ also becomes a dominant
      channel if the channel is open.
\item The decay channel $H_2\rightarrow t\bar{t}$ overwhelms all other
      decay channels as soon as it is allowed.
\item If the CP phase $\Phi$ vanishes, the channel $H_2\rightarrow 
      H_1 H_1$ is prohibited, reflecting that the CP--odd 
      state cannot decay into two identical CP--even states when the 
      system is CP--invariant. However, the contribution of this decay
      mode becomes sizable or suppressed as the CP phase $\Phi$ varies
      from zero.
\item The decay channels $H_2\rightarrow W^+W^-$ and $H_2\rightarrow ZZ$
      are prohibited for the vanishing CP phase. But, they along with 
      the channel $H_2\rightarrow H_1 H_1$ can be also
      sizable and become the most dominant channel for some values of 
      the CP phase $\Phi$.
\end{itemize}
These interesting features are partly because of 
a significant mixing between CP--even and CP--odd states. As shown 
clearly in Fig.~5 of the work by Pilaftsis and Wanger \cite{PW}, 
the couplings of $H_2$ to $H_1 Z$ and, in particular, to $ZZ$ 
increases with the phase $\Phi$ up to 90$^0$ so that they 
clearly enhance the decay channels $H_2\rightarrow H_1 Z$.
On the other hand, the dominance of the decay channel
$H_2\rightarrow H_1 H_1$ for a large mass range of the Higgs boson
$H_2$ is mainly due to a large enhancement of the coupling strength
of the $H_2H_1H_1$ vertex with a non--trivial phase. This distinct
and unique 
feature may give a strong hint of the existence of a non--vanishing
CP phase in the stop or sbottom sector.

When the CP phase is zero, the heaviest Higgs boson is the heavier 
CP--even  state so that the decay channels $H_3\rightarrow H_1 H_1$ and
$H_3\rightarrow VV$ ($V=W$ of $Z$) are among the dominant channels.
This fact is clearly shown in Figure~5 for the partial branching ratios
of various $H_3$ decay channels with respect to the mass $m_{H_3}$. 
In every frame of Figure~5 the upper solid line is for the
sum of all available fermionic decay channels, two dotted lines for
the decay channels into two gauge bosons, the dot--dashed line for the
decay channel $H_3\rightarrow H_1 Z$, 
the dashed line for the decay channel $H_3\rightarrow H_1 H_1$
and finally the lower solid
line for the two--gluon mode. These decays of the heaviest Higgs boson
$H_3$ also exhibit several interesting features:
\begin{itemize}
\item Before the $t\bar{t}$ channel is open, the channels 
      $H_3\rightarrow H_1 H_1$ and $H_3\rightarrow VV$ are dominant,
      while they are overwhelmed by the $t\bar{t}$ mode as soon
      as the latter mode is available. 
      We note the similar behavior of ${\cal B}(H_3\rightarrow VV)$
      and ${\cal B}(H_2 \rightarrow H_1 Z)$ presented in Figure 4,
      a natural consequence from the relation (\ref{eq:relation})
      among the couplings of the neutral Higgs bosons to gauge bosons.
\item The channel $H_3\rightarrow H_2 H_1$ is always prohibited, which
      reflects that three Higgs mass eigenstates are originated from
      two CP--even states and one CP--odd state, and there do not exist
      any interactions between two different CP--even states and one
      CP--odd state. 
\item The channel $H_3\rightarrow H_2 Z$ is not kinematically allowed
      for the whole mass range
      even though there exists the coupling for the decay mode.
\item The decay channel $H_3\rightarrow H_1 Z$ is prohibited for 
      the vanishing CP phase, but the contribution of the decay channel
      to the total width becomes sizable as the CP phase varies from
      zero. Moreover, this mode overwhelms the other decay modes
      near the lower tail of the allowed $H_3$ mass for some nontrivial
      values of the CP phase. We note again the similar  behavior of 
      ${\cal B}(H_3\rightarrow H_1 Z)$ and ${\cal B}(H_2 \rightarrow V V)$ 
      presented in Figure 4.
\end{itemize}
All these interesting features can be understood by noting that the
CP--even and CP--odd compositions of the heavy Higgs mass eigenstates
are strongly dependent on the value of th CP phase $\Phi$. 

To recapitulate, the mass spectrum and the branching fractions of the
lightest Higgs boson $H_1$ are not so sensitive to the CP--violating phase, 
but those of the heavy Higgs states are very sensitive to the phase. 
It means that the CP phase may cause in most cases a larger mixing of 
the CP--odd state with the heavy CP--even state than the light CP--even 
state, although the mixing with the light one also may be substantial,
in particular for light charged Higgs boson masses.
{}From our detailed analysis, it is clear that the CP phase will
affect the phenomenology of the heavy Higgs--boson decays very significantly
so that the heavy Higgs boson searches will depend very strongly on
the CP phase as well as the other SUSY parameters.

\section{Measuring CP violation in fermionic Higgs--boson decays}
\label{sec:measure}

In Section~\ref{sec:decay}, we have found that the branching ratios 
of the lightest Higgs boson decays are (almost) independent of the CP 
phase $\Phi$, and we have noted that for a large value of $\tan\beta$ 
the neutral Higgs boson masses and the branching ratios of the other 
neutral Higgs boson decays also become insensitive to the CP phase. 
However, the CP--violation effects caused by the loop corrections of 
soft CP--violating Yukawa couplings can be directly measured by an 
azimuthal CP--odd decay 
asymmetry in the decays $H_i\rightarrow \tau^+\tau^-$ and/or $H_i\rightarrow
t\bar{t}$ \cite{CKCK} reflecting the correlations of two transversely 
polarized fermions. In this section, we estimate the size of the 
azimuthal CP--odd  
asymmetry in the MSSM with explicit CP violation through radiative 
corrections.

For the sake of discussion, we calculate the helicity amplitude 
explicitly for the Higgs boson decays $H_i\rightarrow f\bar{f}$ by 
use of the so--called 2--component spinor technique \cite{HZ}. 
The result is written as
\begin{eqnarray}
D_{\sigma\bar{\sigma}}=\frac{gm_fm_{H_i}}{2m_WR^f_\beta}
  \left[\sigma\beta_f v^i_f + i\bar{R}^f_\beta a^i_f\right]
  \delta_{\sigma\bar{\sigma}}\,,
\end{eqnarray}
where $\sigma$ and $\bar{\sigma}$ are for the helicity of the fermions
$f$ and $\bar{f}$, respectively, with $f=\tau$ or $t$. 
It is now clear that the helicity amplitude can be complex when both 
the scalar and pseudoscalar parts co-exist and the
interference, signaling CP violation, can be extracted by adjusting the
polarization of the final fermions. As a result, we obtain the following
CP--odd asymmetry in terms of $v^i_f$, $a^i_f$, and $\bar{R}^f_\beta$:
\begin{eqnarray}
{\cal A}^i_{\rm CP}\equiv\frac{2\,{\cal I}\left(D_{--}\, D^*_{++}\right)}{
                  |D_{++}|^2+|D_{--}|^2}=\frac{2\beta_f \bar{R}^f_\beta 
                   v^i_f a^i_f}{|\beta_f v^i_f|^2
                  +|\bar{R}^f_\beta a^i_f|^2}\,.
\end{eqnarray}

In order to observe the CP violating spin--correlation of the top
quarks or $\tau$--leptons, we need to look at the correlations
among their decay products. The analysis power for the spins depends on
each decay mode \cite{AAS}. 
For example, if the top quark decays semi--leptonically,
the spin analysis power is unity, but if it decays hadronically, the 
analysis power is about $0.4$. Similarly, if the $\tau$ decays 
semi--leptonically, the analysis power is $-1/3$. And, the inclusive
hadronic decays of the $\tau$ is known to give the analysis power of
$-0.42$. Certainly, if one can analyze the detailed structure of the
hadronic decays, one can obtain better results. 

Although we can do a more comprehensive numerical analysis for 
evaluating the possibility of detecting CP violation in the fermion
decays, for simplicity we confine ourselves to estimating the size of the
CP--odd asymmetry in the present work. Since the effectiveness of a 
decay channel in measuring CP violation directly depends on the number 
of events as well as the size of the CP--odd asymmetry, it is useful
to define an effective CP--odd asymmetry
as 
\begin{eqnarray}
\hat{\cal A}^i_{\rm CP}={\cal A}^i_{\rm CP}\, 
                      \sqrt{{\cal B}(H_i\rightarrow f\bar{f})}\,.
\end{eqnarray}
With this new effective CP--odd quantity, we can easily estimate the number 
of Higgs bosons required to see the CP--violation effect directly
at 1--$\sigma$ level. The number before including the detection efficiency
and the polarization analysis powers is simply given by
\begin{eqnarray}
N_{H_i}=\frac{1}{(\hat{\cal A}^i_{\rm CP})^2}\,.
\end{eqnarray}
%
For a more realistic estimate, we refer to the work by Atwood and
Soni of Ref.~\cite{CKCK}.  

The CP--odd asymmetry is strongly dependent on the value of $\tan\beta$.
In particular, the $\tau^+\tau^-$ mode is expected to be greatly 
enhanced as $\tan\beta$ increases. In this light, it is interesting
to consider a large $\tan\beta$ case as well, so we consider two
values of $\tan\beta=3$ and $30$ for comparison.
The dependence of the CP--odd asymmetry on the charged Higgs boson
mass $m_{H^\pm}$ for several values of the CP phase in the
decays $H_{1,2,3}\rightarrow \tau^+\tau^-$ is present in Figures~6, 7, and
8, respectively, for two values of $\tan\beta$; the solid line
is for $\tan\beta=3$ and the dashed line for $\tan\beta=30$.
We find that the CP--odd asymmetry in the decay $H_1\rightarrow 
\tau^+\tau^-$ is very sensitive to the charge Higgs boson mass $m_{H^\pm}$  
and can be significant near the lower tail of its allowed mass range for 
non--trivial values of the CP phase. Otherwise, the CP--odd asymmetry 
decreases as $m_{H^\pm}$ increases. This is due to the suppression of the 
mixing matrix element 
$O_{23}$ representing the composition of the CP--odd state in the lightest
Higgs boson. Similarly, near the lower tail of the allowed charged
Higgs boson mass, the CP--odd asymmetry for $H_2\rightarrow \tau^+\tau^-$ 
also is very sensitive to the charged Higgs boson mass. Otherwise, the 
CP--odd asymmetries remain constant instead of decreasing for the
whole range of the charged Higgs boson mass, but they oscillate between
-1 and 1 as the CP phase varies. The CP--odd asymmetry for the
decay $H_3\rightarrow \tau^+\tau^-$ is (almost) independent of the 
charged Higgs boson mass, but it is also very sensitive to the CP--phase.
On the other hand, the large value of the top--quark mass
prohibits the lightest Higgs boson from decaying into a top--quark
pair. So, we consider the decays $H_{2,3}\rightarrow t\bar{t}$, 
of which the dependence on the charged Higgs boson mass $m_{H^\pm}$ for
several values of the CP phase is presented in Figures~9 and 10.
Near the lower tail of the allowed charged Higgs boson mass,
the CP--odd asymmetry for $H_2\rightarrow t\bar{t}$ is very sensitive to 
the charged Higgs boson mass as that for $H_2\rightarrow \tau^+\tau^-$. 
Otherwise, the CP--odd asymmetry remains constant for the
whole range of the charged Higgs boson mass, but it oscillates between
-1 and 1 as the CP phase varies. the CP--odd asymmetry for the
decay $H_3\rightarrow t\bar{t}$ are almost independent of the charged 
Higgs boson mass, but it is also sensitive to the CP phase.

\section{Conclusions}
\label{sec:conclude}

In this paper, we have performed a systematic study of the neutral 
Higgs boson decays in the MSSM with explicit CP violation, which is induced 
through loop corrections involving trilinear CP--violating couplings
of the neutral Higgs bosons to scalar top and bottom quarks and can be 
sizable.
Certainly, our analysis is far from complete in the sense that any 
below--threshold three--body decay channels and comprehensive QCD
corrections, which may be significant, are not included. Therefore, a
more complete work is needed and it is to be reported elsewhere.

The analysis on the branching ratios of the neutral Higgs boson
decays shows that the decay pattern of the lightest
neutral Higgs boson is almost independent of the CP phase, but 
the decay patterns of two other neutral Higgs bosons are
very sensitive to the phase. Since the explicit CP violation is 
realized through the scalar--pseudoscalar mixing between the CP--odd
Higgs state and, mainly, the heavy CP--even Higgs state, 
the decay patterns of two mass eigenstates are determined by the
size of the mixing. Moreover, the dependence of each decay mode on
the CP phase is different. Therefore, it is naturally expected that 
as the CP phase varies from zero, the branching ratios change a lot. 
In particular, for the intermediate Higgs boson (which is nothing but 
the CP--odd Higgs state in the CP invariant case) the channels $H_2
\rightarrow H_1 H_1$ and $H_2\rightarrow VV$ ($V=W$ or $Z$), 
prohibited for the vanishing CP phase and so unique to the CP--noninvariant
theory, can be sizable and become the
most dominant channels for some nontrivial values of the CP phase.
Furthermore, for the heaviest Higgs state (which is the heavier CP--even
state in the CP invariant limit) the decay channel $H_3\rightarrow H_1 Z$
prohibited for the vanishing CP phase may overwhelm the other possible 
decay channels near the lower tail of the allowed $H_3$ mass range for
some nontrivial values of the CP phase.

We have also presented a numerical analysis of the CP--odd asymmetry
that can be extracted through spin correlations of the fermions
in the fermionic decays of the neutral Higgs
bosons, $H_i\rightarrow \tau^+\tau^-$ and $H_i\rightarrow t\bar{t}$.
The CP--odd asymmetry is very sensitive to the value of $\tan\beta$,
in particular, for the decay $H_1\rightarrow \tau^+\tau^-$. For a large
value of $\tan\beta$, the branching ratio for the decay channel is
enhanced and moreover the CP--odd asymmetry becomes significant for
small Higgs boson masses. However, we have found that the CP--odd 
asymmetry for the Higgs boson $H_1$ decreases as the charged Higgs 
boson mass $m_{H^\pm}$ increases increases
due to the suppression of $O_{23}$. Except the lower tail of the allowed
charged Higgs boson mass, the CP--odd asymmetries for both 
$H_{2,3}\rightarrow \tau^+\tau^-$ and $H_{2,3}\rightarrow t\bar{t}$
remain constant for the charged Higgs boson mass, but 
On the whole, the CP--odd asymmetry can
be saturated to be 1 so that there is a great chance to see CP violation
due to radiative corrections directly through the spin correlations
of the tau leptons and/ot top--quarks in the fermionic decays of any 
neutral Higgs boson.

To conclude, the analysis presented in this paper clearly demonstrates
that the explicit CP violation in the MSSM Higgs sector may modify 
the Higgs boson decays significantly so that it will have a great impact
on Higgs searches, and give a great chance to measure the effects
directly through the CP--odd spin correlations in the
decays $H_i\rightarrow \tau^+\tau^-$ and/or $H_i\rightarrow t\bar{t}$.

\section*{Acknowledgments}

The authors thank D.A.~Demir for useful comments.
S.Y.C wishes to acknowledge the financial support of 1997-sughak program
of Korean Research Foundation and he also would like to thank F. Halzen and 
the Physics Department, University of Wisconsin - Madison where part of the
work has been carried out.

\setcounter{equation}{0}
\renewcommand{\theequation}{A\arabic{equation}}

\section*{Appendix}

\def\beq{\begin{equation}}
\def\eeq{\end{equation}}
\def\beqar{\begin{eqnarray}}
\def\eeqar{\end{eqnarray}}
\def\Re#1{{\cal R}\left({#1}\right)}
\def\Im#1{{\cal I}\left({#1}\right)}
\def\sb#1{{s_\beta^{#1}}}
\def\cb#1{{c_\beta^{#1}}}
\def\tb#1{{t_\beta^{#1}}}
\def\l#1{\lambda_{#1}}
\def\half{\frac{1}{2}}
\def\e1{{\rm e}^{i\xi}}

In this Appendix, we present the explicit form of the coefficients 
$b_\alpha$ and the fully--symmetric trilinear Higgs self-couplings 
$a_{\alpha\beta\gamma}$, $(\alpha=1,2,3)$ in terms of the quartic 
couplings $\lambda_i$ ($i=1$ to $8$), which are needed
for the Higgs boson decays into a charged Higgs--boson pair and
a neutral Higgs--boson pair, respectively. The coefficients describing the
couplings of the neutral Higgs boson to a charged Higgs-boson pair
in the weak eigenstates are given by 
\begin{eqnarray}
b_1 &=& 2 \sb{} \cb{} \Im{\l{5}{\rm e}^{2i\xi}}
       -\sb{2} \Im{\l{6}\e1}-\cb{2} \Im{\l{7}\e1} \, , \nonumber \\ 
b_2 &=& 2 \sb{2} \cb{} \l{1} +\cb{3}\l{3}-\sb{2}\cb{}\l{4}
       -2\sb{2}\cb{}\Re{\l{5}{\rm e}^{2i\xi}}+\sb{}(\sb{2}-2\cb{2})\Re{\l{6}\e1}
       +\sb{}\cb{2}\Re{\l{7}\e1} \, ,
        \nonumber \\ 
b_3 &=& 2 \cb{2} \sb{} \l{2} +\sb{3}\l{3}-\cb{2}\sb{}\l{4} 
       -2\cb{2}\sb{}\Re{\l{5}{\rm e}^{2i\xi}}+\cb{}\sb{2}\Re{\l{6}\e1}
       +\cb{}(\cb{2}-2\sb{2})\Re{\l{7}\e1}  \, ,
\end{eqnarray}
and the symmetric coefficients $a_{\alpha\beta\gamma}$ describing the
self--interactions of three neutral Higgs bosons by
\begin{eqnarray}
a_{111} &=& \sb{}\cb{} \Im{\l{5}{\rm e}^{2i\xi}}-\half\sb{2}\Im{\l{6}\e1}
            -\half\cb{2}\Im{\l{7}\e1} \, ,
            \nonumber \\
a_{112} &=&\frac{1}{3}\left[\sb{2}\cb{}\l{1}+\half\cb{3}(\l{3}
           +\l{4})-\cb{}(1+\sb{2})\Re{\l{5}{\rm e}^{2i\xi}} \right.
            \nonumber \\
&&\left.   +\half\sb{}(\sb{2}-2\cb{2})\Re{\l{6}\e1}
           +\half\cb{2}\sb{}\Re{\l{7}\e1} \right] \, ,
            \nonumber \\
a_{113} &=&\frac{1}{3}\left[\cb{2}\sb{}\l{2}+\half\sb{3}(\l{3}
           +\l{4})-\sb{}(1+\cb{2})\Re{\l{5}{\rm e}^{2i\xi}} \right.
            \nonumber \\
&&\left.   +\half\sb{2}\cb{}\Re{\l{6}\e1}+\half\cb{}(\cb{2}
           -2\sb{2})\Re{\l{7}\e1}\right] \, ,
            \nonumber \\
a_{122}&=&\frac{1}{3}\left[-\sb{}\cb{}\Im{\l{5}{\rm e}^{2i\xi}}
           -\half(1+2\cb{2})\Im{\l{6}\e1}\right] \, ,
            \nonumber \\
a_{123}&=&\frac{1}{6}\left[-2\Im{\l{5}{\rm e}^{2i\xi}}-\cb{}\sb{}(\Im{\l{6}\e1}
           +\Im{\l{7}\e1})\right] \, ,
            \nonumber \\
a_{133}&=&\frac{1}{3}\left[-\sb{}\cb{}\Im{\l{5}{\rm e}^{2i\xi}}
           -\half(1+2\sb{2})\Im{\l{7}\e1}\right] \, ,
            \nonumber \\
a_{222}&=& \cb{}\l{1}+\half\sb{}\Re{\l{6}\e1} \, ,
            \nonumber \\
a_{223}&=&\frac{1}{3}\left[\half\sb{}(\l{3}+\l{4})+\sb{}\Re{\l{5}{\rm e}^{2i\xi}}
           +\frac{3}{2}\cb{}\Re{\l{6}\e1}\right] \, ,
           \nonumber \\
a_{233}&=&\frac{1}{3}\left[\half\cb{}(\l{3}+\l{4})+\cb{}\Re{\l{5}{\rm e}^{2i\xi}}
           +\frac{3}{2}\sb{}\Re{\l{7}\e1}\right] \, ,
           \nonumber \\
a_{333}&=& \sb{}\l{2}+\half\cb{}\Re{\l{7}\e1} \, .
\end{eqnarray}
The remaining 17 coefficients of $a_{\alpha\beta\gamma}$ can be obtained 
by the symmetry properties $a_{\alpha\beta\gamma}=a_{\beta\gamma\alpha}=
a_{\gamma\alpha\beta}=a_{\alpha\gamma\beta}=a_{\gamma\beta\alpha}=
a_{\beta\alpha\gamma}$. We note  that $b_1$, $a_{111}$, $a_{122}$,
$a_{123}$, and $a_{133}$ come only from the imaginary parts of one-loop 
corrections to the effective Higgs potential so that these coefficients 
vanish in the CP--invariant theory.

\vskip 1.5cm

\begin{figure}
 \begin{center}
\hbox to\textwidth{\hss\epsfig{file=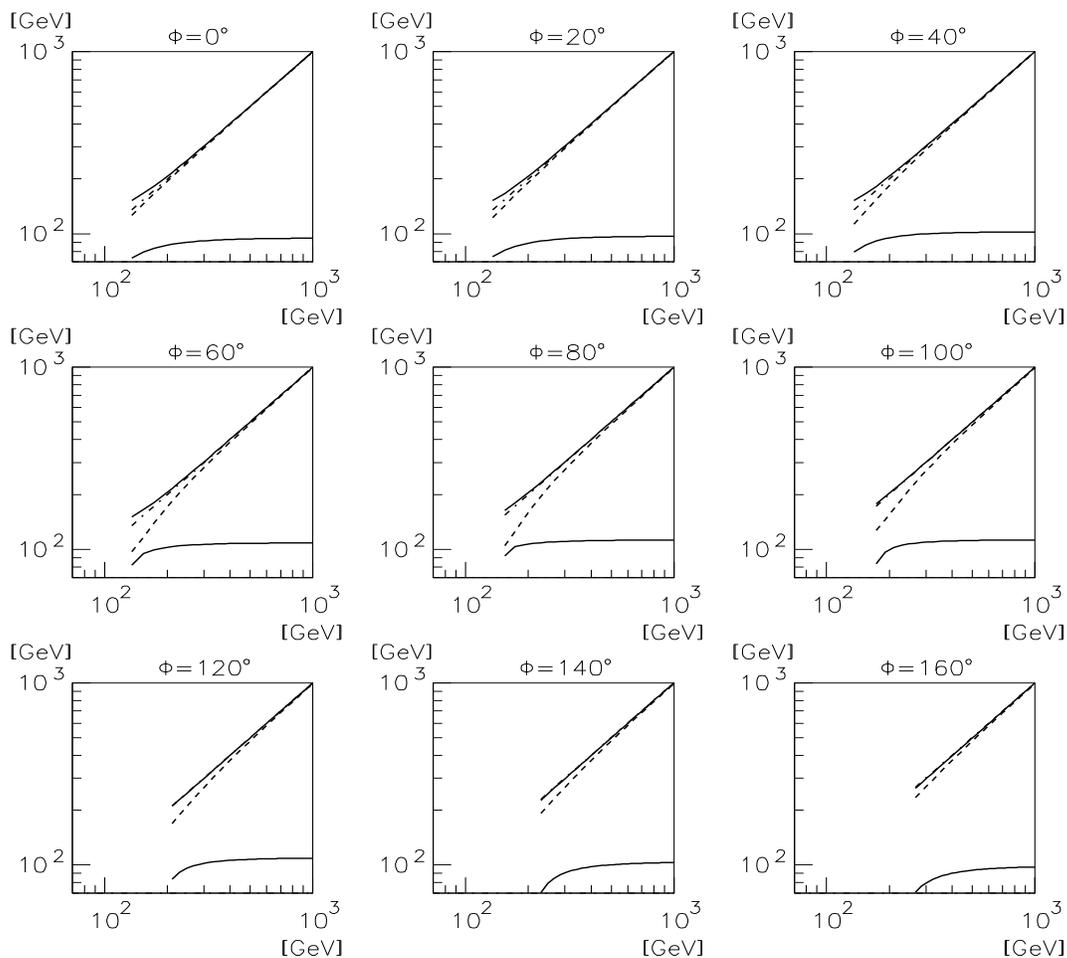,width=16cm,height=14cm}\hss}
 \end{center}
\caption{Higgs--boson masses with respect to the charged Higgs mass 
         $m_{H^\pm}$ for different values of the CP phase $\Phi$ with 
         the fixed values for other SUSY parameters in 
         Eq.~(\ref{eq:parameter}). The value of $\tan\beta$ is taken to 
         be 3 and an artificial experimental condition $m_{H_1}\geq 70$ 
         GeV is imposed. The lower solid line is for the lightest Higgs boson 
         $H_1$, the upper solid line for the heaviest Higgs boson $H_3$, 
         and the dashed line for the intermediate Higgs boson $H_2$. 
         For reference, the charged Higgs mass is presented with the 
         dot--dashed line.}
\label{fig1}
\end{figure}

\vskip 1.5cm

\begin{figure}
 \begin{center}
\hbox to\textwidth{\hss\epsfig{file=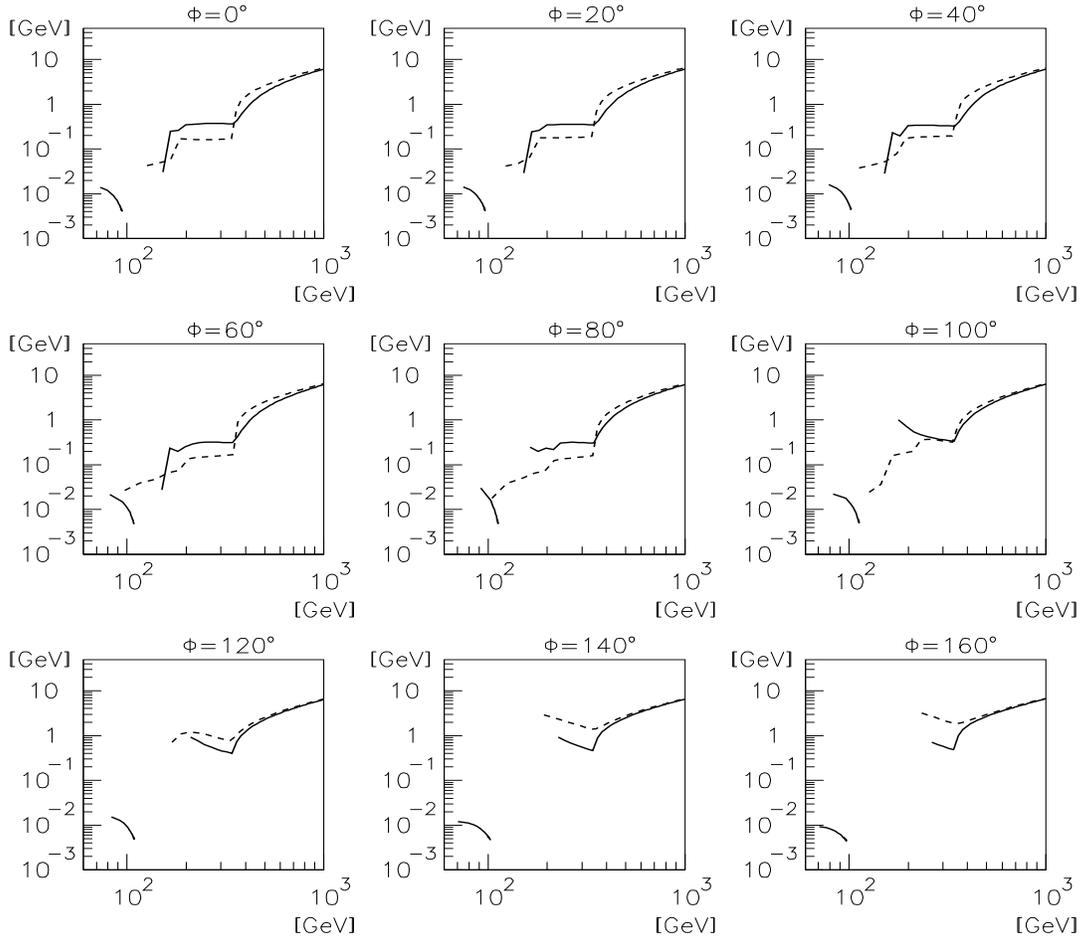,width=16cm,height=14cm}\hss}
 \end{center}
\caption{Total decay widths for the three neutral Higgs bosons with respect
         to each Higgs--boson mass for several values of the CP phase
         $\Phi$. The left (right) solid 
         line is for the lightest (heaviest) Higgs boson and the dashed 
         line for the intermediate Higgs boson.}
\label{fig2}
\end{figure}

\vskip 1.5cm

\begin{figure}
 \begin{center}
\hbox to\textwidth{\hss\epsfig{file=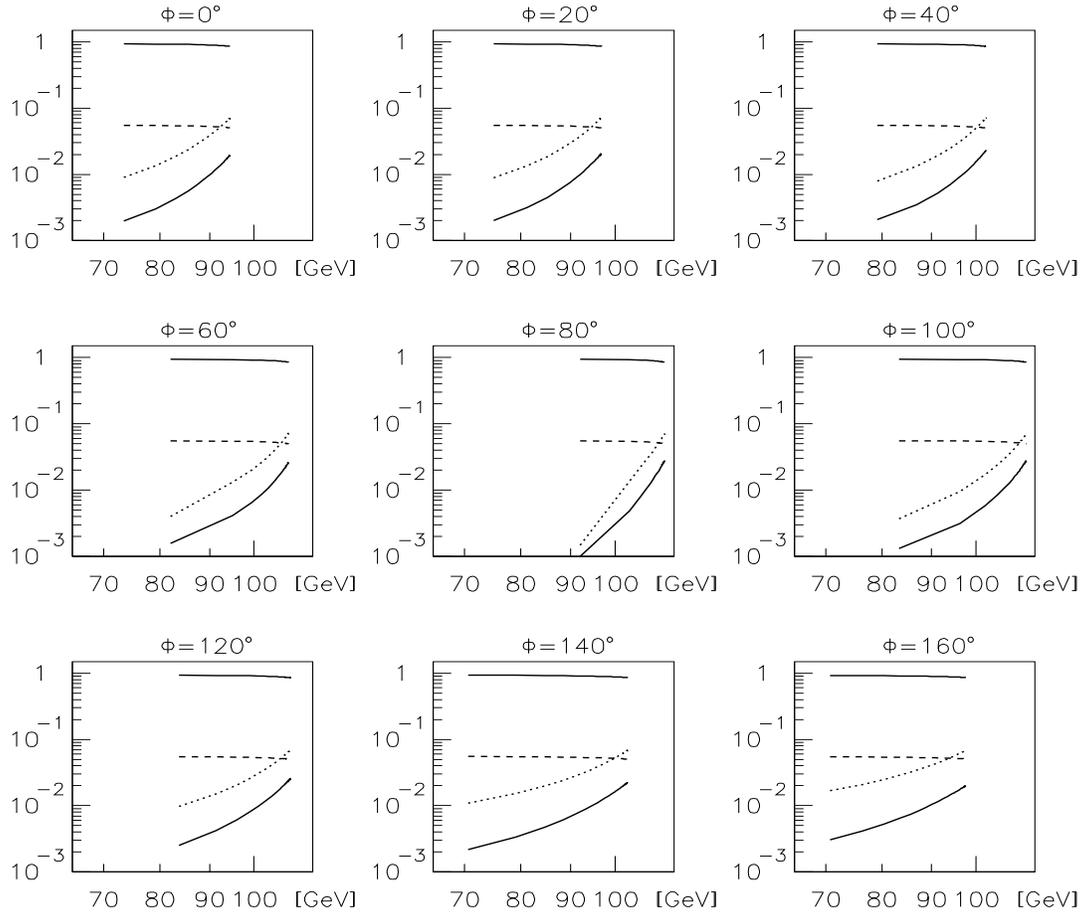,width=16cm,height=14cm}\hss}
 \end{center}
\caption{Partial branching ratios of the lightest Higgs boson decays 
         with respect to the mass $m_{H_1}$ for sevaral 
         values of the CP phase $\Phi$. The upper solid line is for the
         channel $H_1\rightarrow b\bar{b}$, the dashed line for the
         channel $H_1\rightarrow \tau^+\tau^-$, the dotted line for
         the channel $H_1\rightarrow c\bar{c}$ and the lower solid line
         for the two--gluon mode $H_1\rightarrow gg$.}
\label{fig3}
\end{figure}

\vskip 1.5cm

\begin{figure}
 \begin{center}
\hbox to\textwidth{\hss\epsfig{file=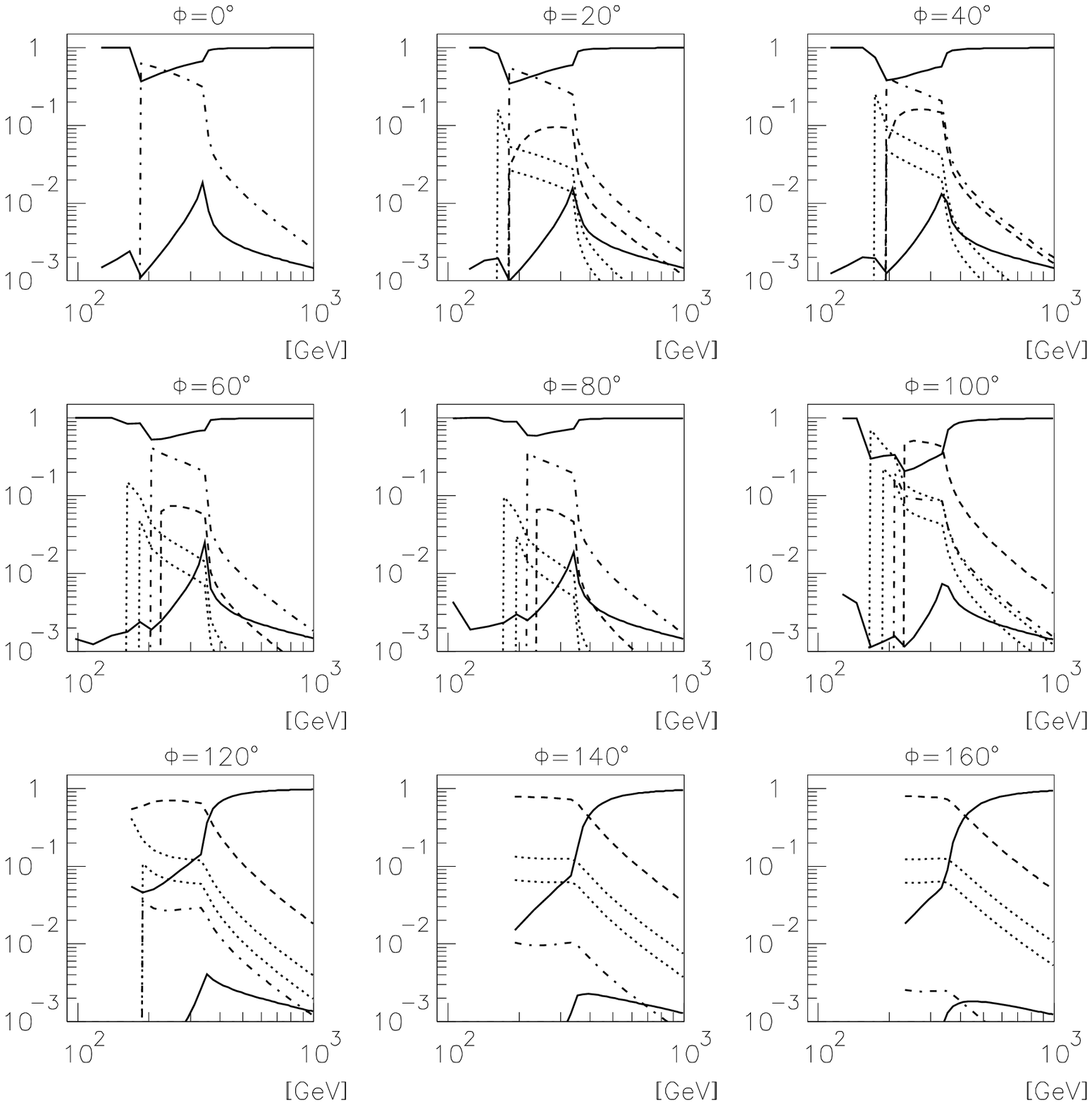,width=16cm,height=14cm}\hss}
 \end{center}
\caption{Partial branching ratios for the $H_2$ decays with respect to the
         mass $m_{H_2}$ for several values of 
         the CP phase $\Phi$. The upper solid line in each figure frame
         is for  the sum of the branching ratios of the fermionic decay 
         modes. The dot--dashed line is for the decay channel 
         $H_2\rightarrow H_1 Z$, two dotted lines for the channels 
         $H_2\rightarrow W^+W^-$ (upper line) and $ZZ$ (lower line),
         respectively, and the dashed line for the decay channel 
         $H_2\rightarrow H_1 H_1$. Finally, the lower solid line is 
         for the two--gluon channel.}
\label{fig4}
\end{figure}

\vskip 1.5cm

\begin{figure}
 \begin{center}
\hbox to\textwidth{\hss\epsfig{file=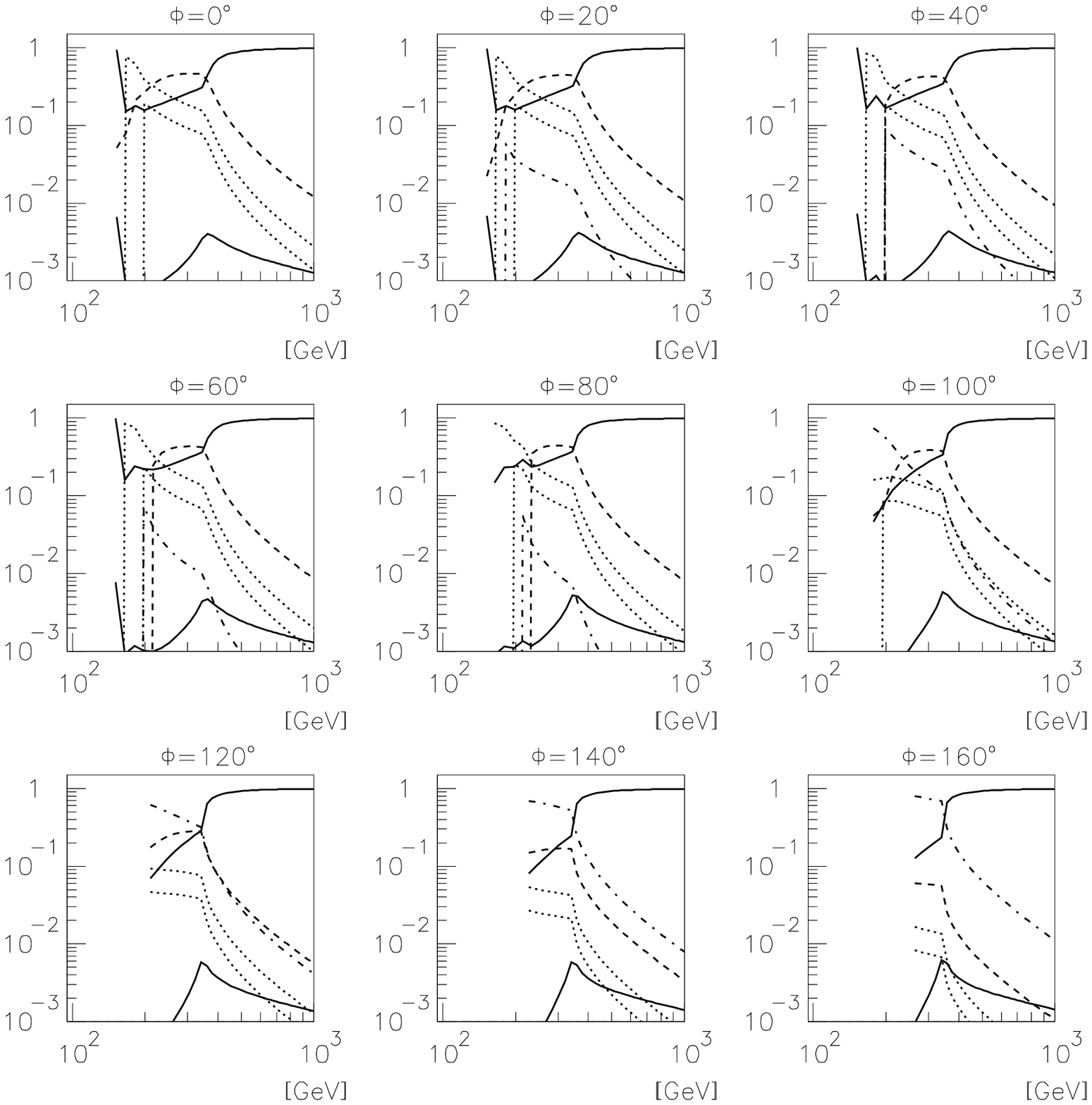,width=16cm,height=14cm}\hss}
 \end{center}
\caption{Partial branching ratios for the $H_3$ decays with respect to the
         mass $m_{H_3}$ for several values of 
         the CP phase $\Phi$. The upper solid line in each figure frame
         is for  the sum of the branching ratios of the fermionic decay 
         modes. The dot--dashed line is for the decay channel 
         $H_3\rightarrow H_1 Z$, two dotted lines for the channels 
         $H_3\rightarrow W^+W^-$ (upper line) and $ZZ$ (lower line),
         respectively, and the dashed line for the decay channel 
         $H_3\rightarrow H_1 H_1$. Finally, the lower solid line is 
         for the two--gluon channel.}
\label{fig5}
\end{figure}

\vskip 1.5cm

\begin{figure}
 \begin{center}
\hbox to\textwidth{\hss\epsfig{file=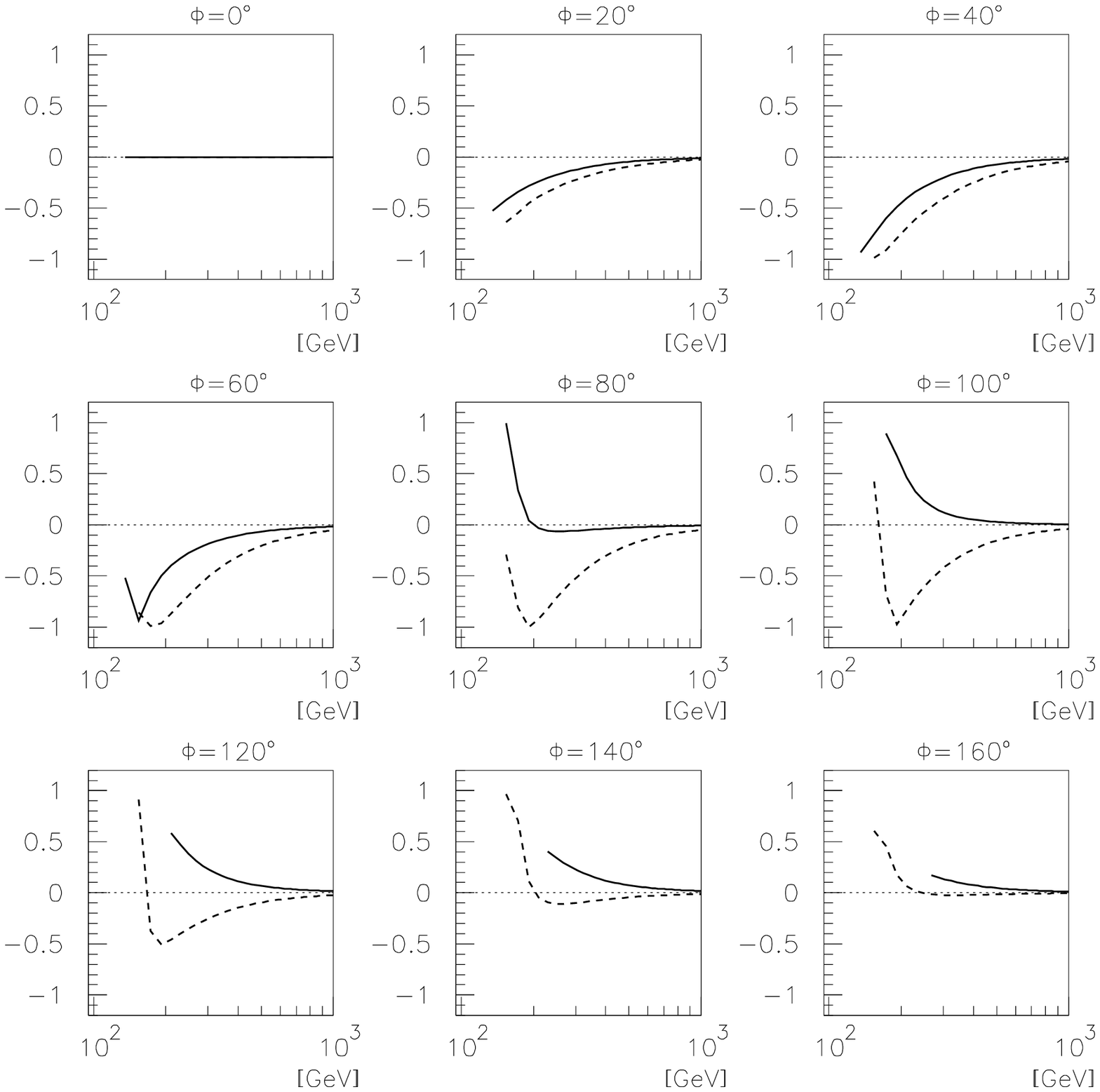,width=16cm,height=14cm}\hss}
 \end{center}
\caption{The CP--odd effective asymmetry $\hat{\cal A}^i_{\rm CP}$ with 
         respect to the charged Higgs boson mass 
         $m_{H^\pm}$ for several values of the CP phase $\Phi$ in the
         decay $H_1\rightarrow \tau^+\tau^-$. The solid line in each 
         frame is for $\tan\beta=3$ and the dashed line for 
         $\tan\beta=30$.}
\label{fig6}
\end{figure}

\vskip 1.5cm

\begin{figure}
 \begin{center}
\hbox to\textwidth{\hss\epsfig{file=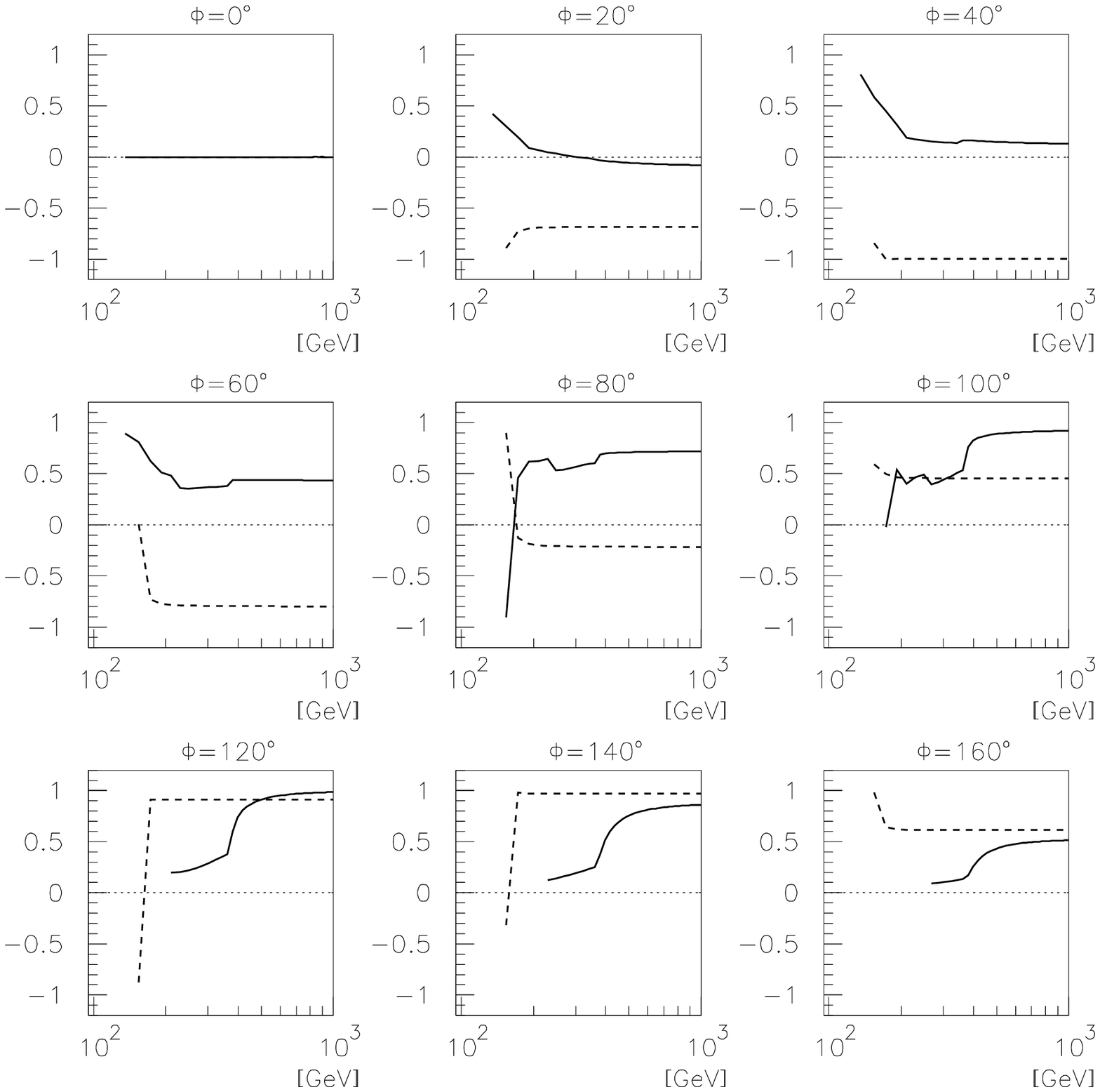,width=16cm,height=14cm}\hss}
 \end{center}
\caption{The CP--odd effective asymmetry $\hat{\cal A}^i_{\rm CP}$ with 
         respect to the charged Higgs boson mass 
         $m_{H^\pm}$ for several values of the CP phase $\Phi$ in the
         decay $H_2\rightarrow \tau^+\tau^-$. The solid line in each 
         frame is for $\tan\beta=3$ and the dashed line for 
         $\tan\beta=30$.}
\label{fig7}
\end{figure}

\vskip 1.5cm

\begin{figure}
 \begin{center}
\hbox to\textwidth{\hss\epsfig{file=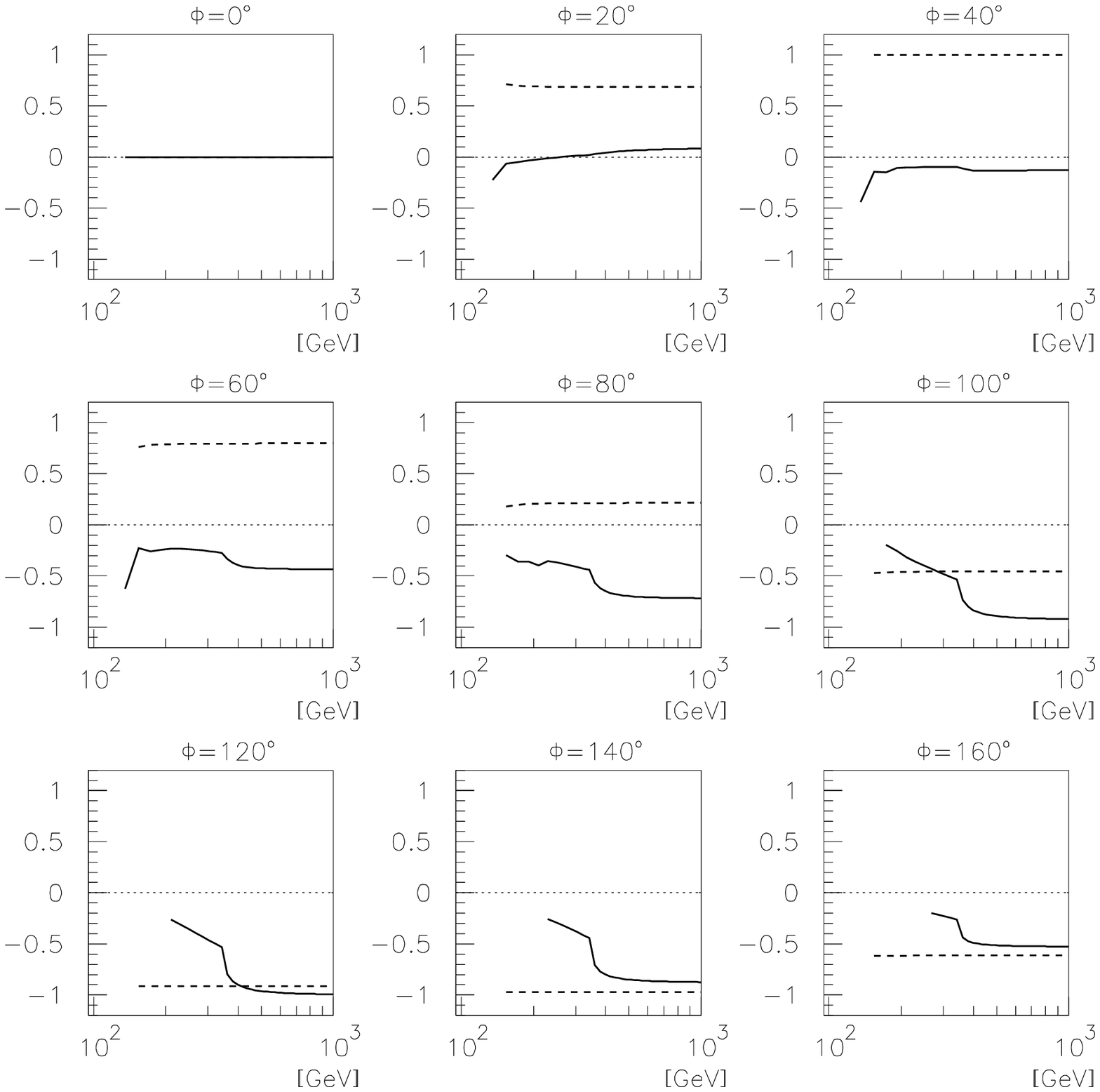,width=16cm,height=14cm}\hss}
 \end{center}
\caption{The CP--odd effective asymmetry $\hat{\cal A}^i_{\rm CP}$ with 
         respect to the charged Higgs boson mass 
         $m_{H^\pm}$ for several values of the CP phase $\Phi$ in the
         decay $H_3\rightarrow \tau^+\tau^-$. The solid line in each 
         frame is for $\tan\beta=3$ and the dashed line for 
         $\tan\beta=30$.}
\label{fig8}
\end{figure}

\vskip 1.5cm

\begin{figure}
 \begin{center}
\hbox to\textwidth{\hss\epsfig{file=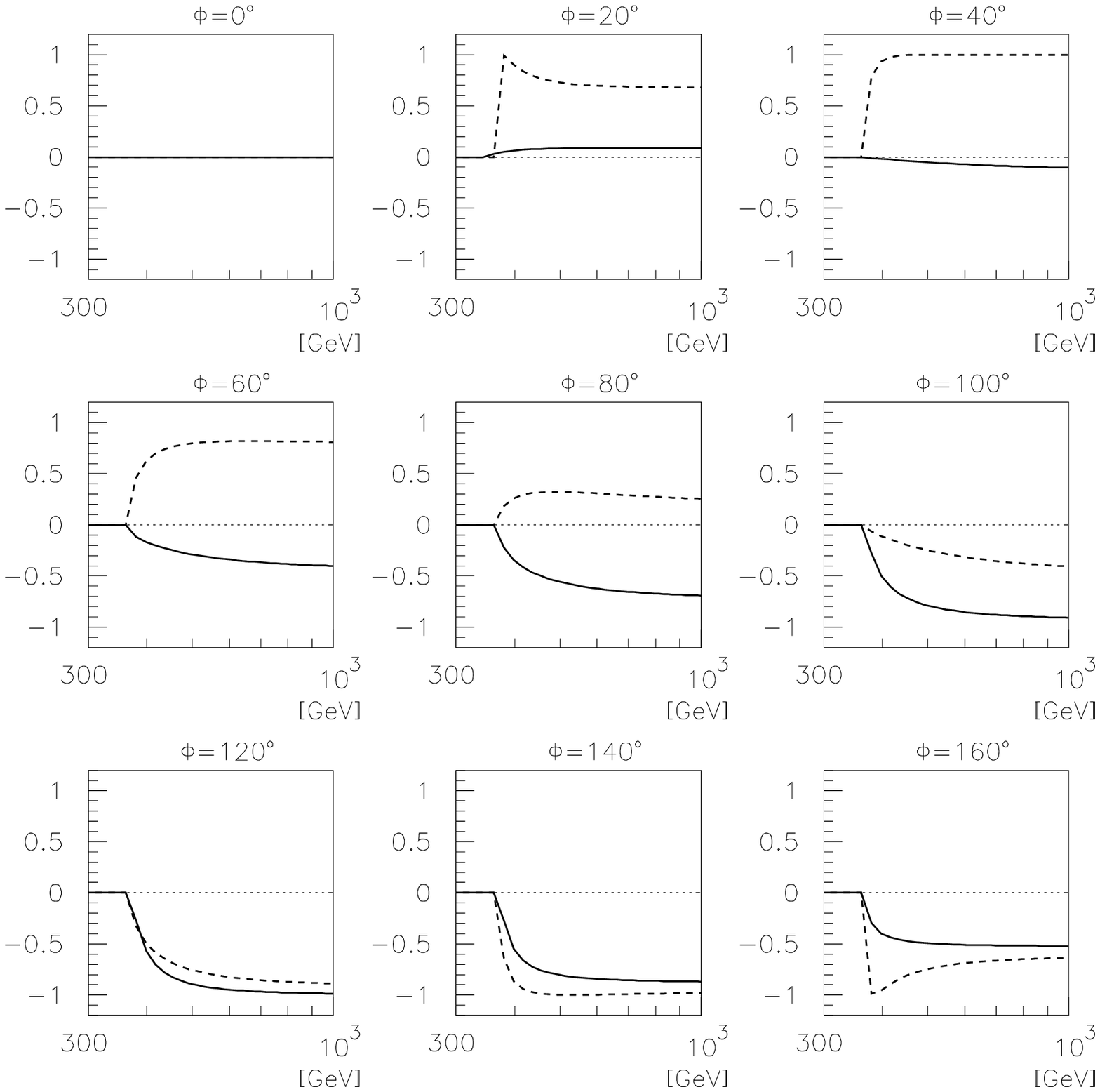,width=16cm,height=14cm}\hss}
 \end{center}
\caption{The CP--odd effective asymmetry $\hat{\cal A}^i_{\rm CP}$
         with respect to the charged Higgs boson mass 
         $m_{H^\pm}$ for several values of the CP phase $\Phi$ in the
         decay $H_2\rightarrow t\bar{t}$. The solid line in each 
         frame is for $\tan\beta=3$ and the dashed line for 
         $\tan\beta=30$.}
\label{fig9}
\end{figure}

\vskip 1.5cm

\begin{figure}
 \begin{center}
\hbox to\textwidth{\hss\epsfig{file=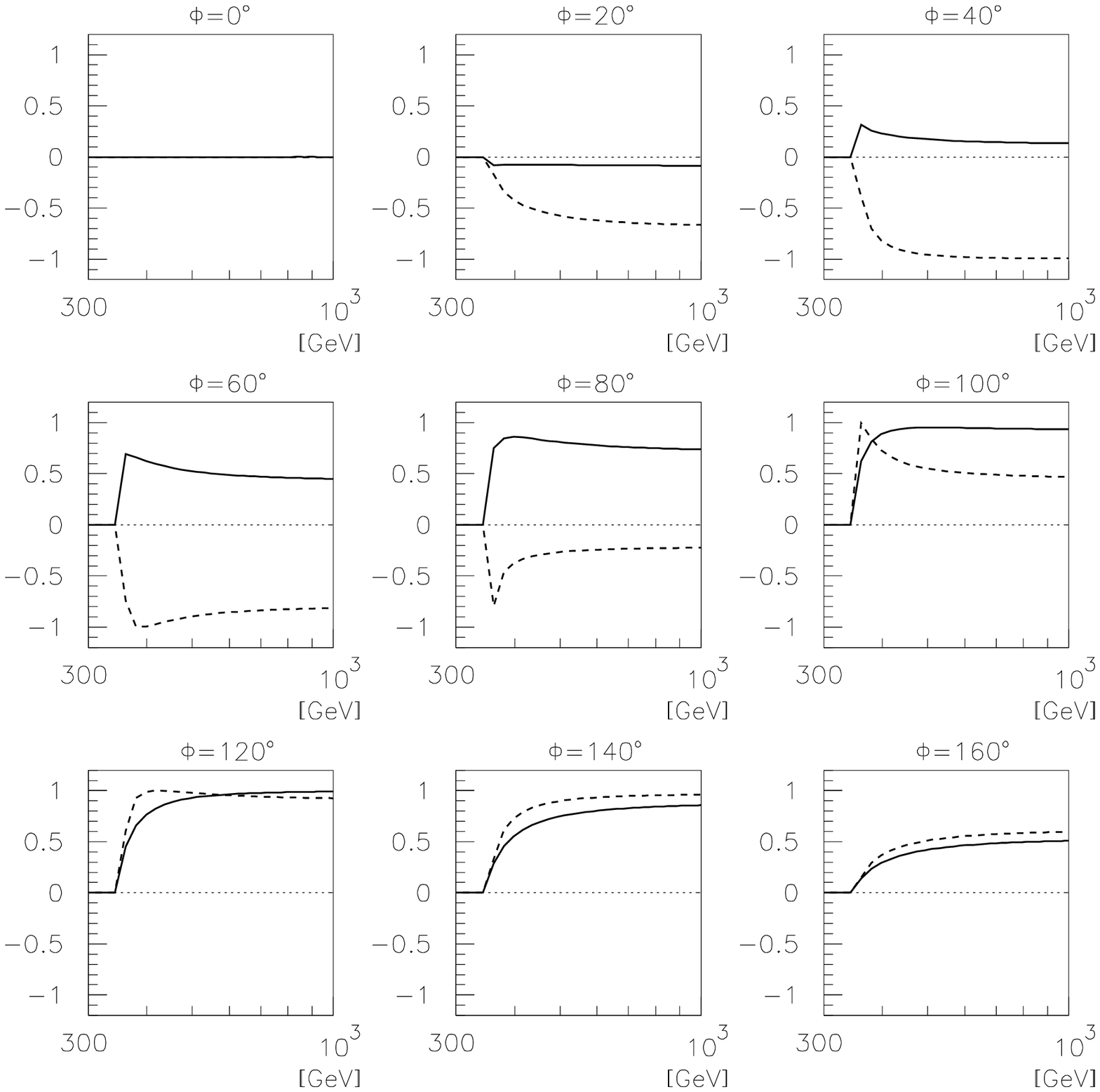,width=16cm,height=14cm}\hss}
 \end{center}
\caption{The CP--odd effective asymmetry $\hat{\cal A}^i_{\rm CP}$ 
         with respect to the charged Higgs boson mass 
         $m_{H^\pm}$ for several values of the CP phase $\Phi$ in the
         decay $H_3\rightarrow t\bar{t}$. The solid line in each 
         frame is for $\tan\beta=3$ and the dashed line for 
         $\tan\beta=30$.}
\label{fig10}
\end{figure}

\vfil\eject


\begin{references}
\bibitem{tanb} A large value of $\tan\beta$ may be worthwhile to be
   considered. However, we have confirmed that in this case the Higgs
   mass spectrum is (almost) independent of the CP phase $\Phi$ and
   all the neutral Higgs decay widths are dominated by the contributions
   from the fermionic Higgs boson decays such as $H_i\rightarrow b\bar{b}$
   and $H_i\rightarrow\tau^+\tau^-$ as well as $H_i\rightarrow t\bar{t}$.
\end{references}
\end{document}